# Coherent effects in crystal collimation

V.M. Biryukov[♦]

*Institute for High Energy Physics, Protvino, 142281, Russia*

**Abstract**

We present theory for coherent effects observed in crystal collimation experiments that is in good quantitative agreement with RHIC and Tevatron data. We show that coherent scattering in a bent crystal strongly amplifies beam diffusion, with an effective radiation length shortened by orders of magnitude compared to amorphous material. This coherent scattering could replace the traditional amorphous scattering in accelerator collimation systems. We predict that crystal collimation for negative particles can be as strong as for positives, unlike with channeling effect. This opens a principle way for efficient crystal steering of negative particles at accelerators. It can be demonstrated with antiproton crystal collimation at the Tevatron. We predict strong effects for the upcoming Tevatron experiment, for protons and antiprotons.

## 1. Introduction

Bent crystal technique is well established for channeling high-energy beams, in particular for beam extraction from accelerators [1]. It was successfully applied from 3 MeV [2] to world highest energy [3], is well understood theoretically [4] and considered as possible instrument for the 7 TeV LHC [5-7]. IHEP Protvino experiments have demonstrated that this technique can be quite efficient [8]: 85% of 70 GeV proton beam is extracted at beam intensity up to $4\times10^{12}$. Much of the IHEP physics program relies on crystal channeled beams used regularly since 1989.

The theory of crystal extraction is based on simulations tracking the particles through a curved crystal lattice and accelerator environment [9,10]. Simulation code CATCH [11] was successfully tested in channeling experiments at CERN SPS [12], Tevatron [13], RHIC [14-16], and IHEP U-70 [4]. Monte Carlo predictions, suggesting a "multipass" mode of crystal extraction where efficiency is dominated by the multiplicity of particle encounters with a short crystal in a ring, lead to the record high efficiency demonstrated at IHEP [8].

It would be promising to apply the bent-crystal technique for a beam halo scraping in the Tevatron [17] and LHC [7] where an order of magnitude reduction is expected in the accelerator-related backgrounds. A bent crystal, serving as a primary element, should coherently bend halo

---

[♦] http://mail.ihep.ru/~biryukov/

particles onto a secondary collimator. Crystal collimation was first proposed and studied in 1989 at IHEP for a 3 TeV UNK [18,19] collider. In 1991 it was studied for 20 TeV SSC [20]. A demonstration experiment on crystal collimation was performed in 1998 at IHEP where a factor-of-2 reduction in the accelerator background was obtained with a bent crystal incorporated into beam cleaning system [21].

## 2. RHIC and Tevatron experiments

Another experiment on crystal collimation was done at the Relativistic Heavy Ion Collider [14-16]. The Yellow ring of the RHIC had a bent Si crystal collimator, 5 mm along the beam. By properly aligning the crystal to the beam halo, particles entering the crystal were deflected away from the beam and intercepted downstream in a copper scraper. RHIC crystal collimator efficiency measured for gold ions as a function of the crystal angle was found in good agreement with simulations with the measured machine optics, as seen in Fig. 1. For the 2003 RHIC run, the theory predicted the efficiency of 32%, and averaging over the data for this run gave the measured efficiency of 26%. The modest figure of efficiency ≈30%, both in theory and experiment, is attributed to the high angular spread of the beam that hits the crystal face as set by machine optics. It is worth to compare this figure of efficiency for gold ions at RHIC to the 40% efficiency achieved with the same (5 mm O-shaped Si) crystal for protons at IHEP in 1998 [22]. Crystal extraction of Pb ions was earlier demonstrated at CERN SPS with efficiency of 4-11% for a long (40 mm) Si crystal [23].

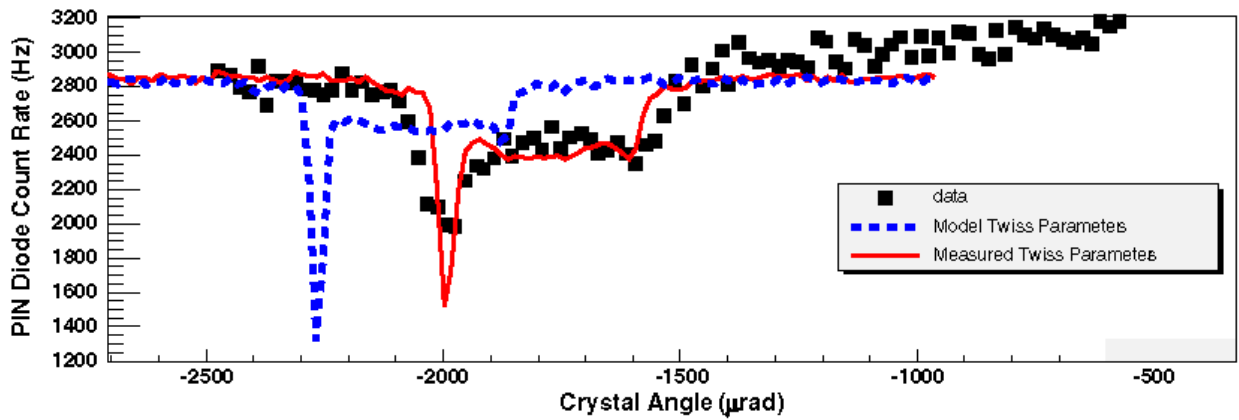

**Figure 1** The rate of nuclear interactions in the crystal measured (dots) and simulated as a function of the crystal orientation at RHIC. Two sets of CATCH simulation shown: with preliminary optics (dash) and with measured optics (solid).

In the analysis of crystal collimation, major attention was paid so far to the peak efficiency of channeling that makes applications so attractive. However, apart from a strong channeling peak, the RHIC collimation experiment has shown another interesting feature. Fig. 1 shows a substantial reduction in the rate of nuclear interactions in the crystal observed over a broad angular range, -2000 to -1500 μrad, corresponding to the crystal bending angle (0.44 mrad). This plateau in the RHIC collimation plots was observed also in simulations [14-16]. While the quantitative agreement was good, the interpretation of the results remained an important issue.

Following the RHIC experiment, a crystal collimator has been installed into the Tevatron, using the same "O-shaped" Si crystal from the RHIC collimation set-up. Remarkably, the data coming from the Tevatron experiment show the effects very similar to RHIC observations, Fig. 2 [24]. The dramatic dip in the local background rate with aligned crystal indicates a very high channeling efficiency, confirming the high expectations for crystal collimation at high energy colliders [7,17]. Moreover, the plateau effect is confirmed and becomes stronger, again in very good quantitative agreement [24] with simulation [25].

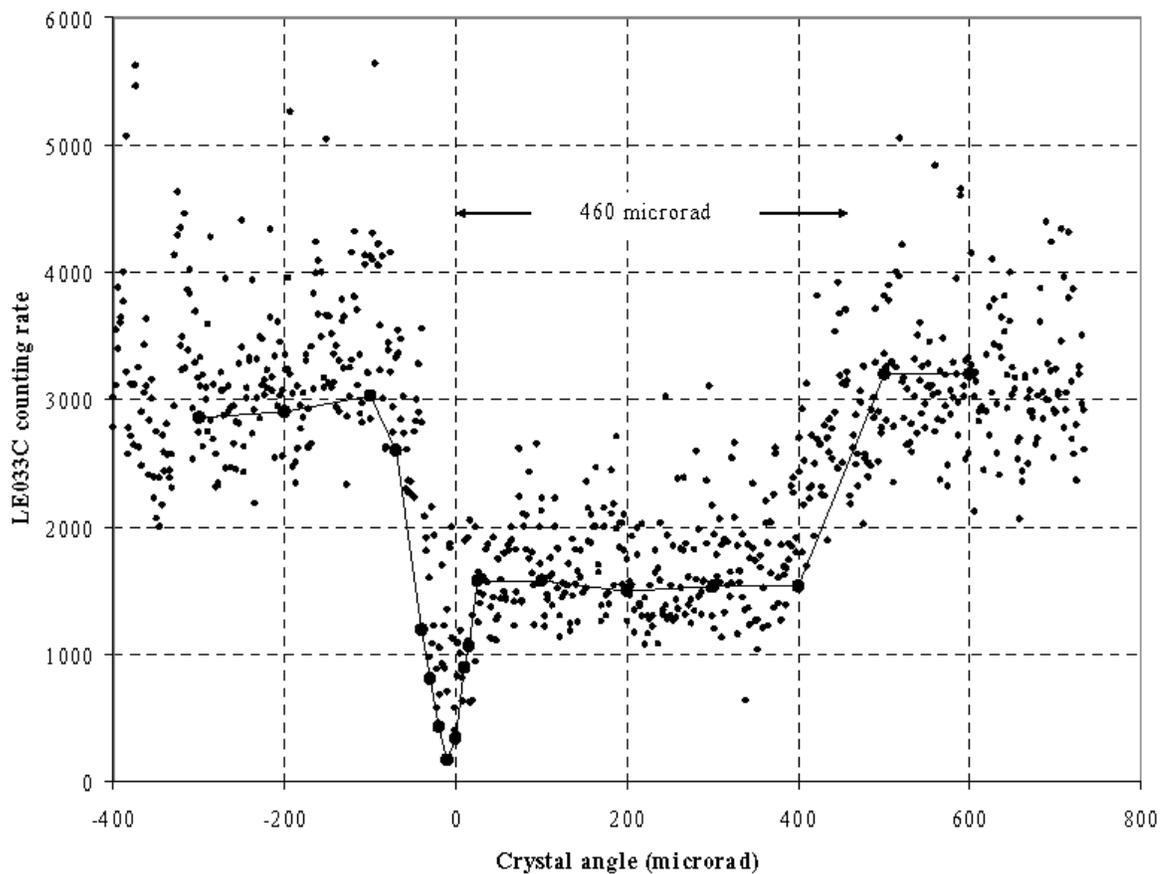

**Figure 2** The rate of nuclear interactions in the crystal measured and simulated (larger circles, CATCH) as a function of the O-shaped crystal orientation at the Tevatron. From ref. [24].

### 3. Simulations and interpretation

Ref. [25] gave explanation of the plateau effect in crystal collimation relating it to the coherent scattering ("reflection") of particles on the field of bent crystal atomic planes. This coherent scattering makes the beam diffusion in accelerator much stronger compared to scattering in amorphous material and affects the particle loss pattern in the accelerator ring.

For the plateau range, every particle becomes tangential to atomic planes somewhere in a bent crystal depth. Two effects in these conditions are known from the physics of channeling and quasi-channeling: "volume capture" (scattering-induced transfers of random particles to channeled states) [26] and "volume reflection" (scattering of random particles off the potential of bent atomic planes) [27].

In order to understand fully the origin of the plateau effect, we simulated RHIC and Tevatron experiments studying the particle dynamics in single and multiple interactions with a bent crystal in different angular ranges in the environment of collimation experiment. At a random crystal orientation, these interactions are similar to those in amorphous media.

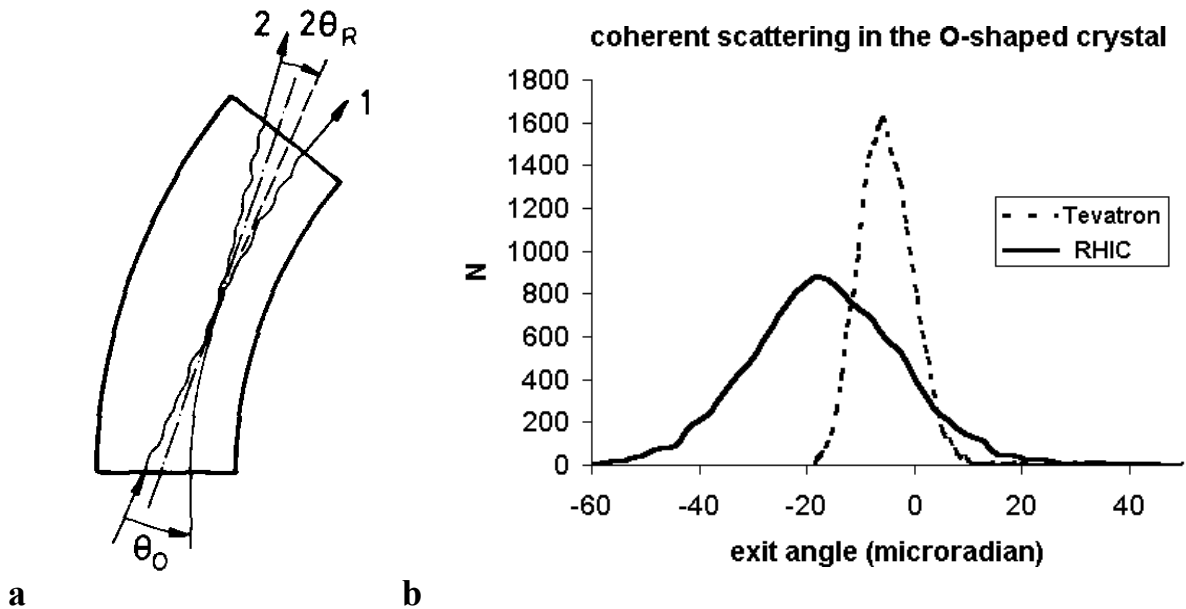

**Figure 3 (a)** Schematic picture of a channeled particle *1* and reflected particle *2*. **(b)** Particle exit angular distributions. Zero angle corresponds to the direction of an incident particle.

Fig. 3(a) shows schematically the case of a particle becoming tangential to atomic planes in a crystal depth [27]. With some probability, such a particle is trapped into channeled state *1*; otherwise, it is scattered ("reflected") in the coherent potential of bent atomic planes. The simulated exit angular distributions of a particle (100 GeV/u gold ion for RHIC and 980 GeV proton for the Tevatron) interacting with the crystal used in the experiments are shown in Fig. 3(b). The distributions show a clear shift in the average exit angle to the side opposite to a crystal bending - this effect is known as volume reflection [27]. The most probable exit angle is about –17 μrad for RHIC and –5.3 μrad for the Tevatron, i.e., 1-2 critical channeling angle (10 μrad for RHIC and 5 μrad for Tevatron). This shift is independent of the incidence angle within the plateau range in Figs. 1-2. Apart from a spectacular shift, the distributions show a substantial broadening. The r.m.s. exit angle of the reflected beam in Fig.3(b) is 22 (RHIC) and 7.2 (Tevatron) μrad, which is much broader than the r.m.s. scattering angle at a random incidence, i.e., 13 μrad for RHIC and 3.3 μrad for the Tevatron. So, a coherent scattering on the potential of bent atomic planes appears a very significant factor for particle dynamics in accelerator.

A small number of particles is volume-captured and channeled over some crystal length, and thus obtains a substantial bending angle. The probability of volume capture into stable channeled states reduces with energy $E$ as $E^{-3/2}$ [28], and is often neglected. However, it can be much higher for unstable states [29] which may cause particle bending of some tens μrad, a very significant value for particle dynamics in accelerator. Typically, in the collimation setups of the considered experiments, a particle deflection of >20-30 μrad at a crystal leads to its immediate loss at a secondary collimator.

For RHIC collimation setup, with the O-shaped crystal at 6σ, the computed probability of particle deflection beyond a 7σ secondary aperture because of a volume capture is 2% per a single encounter with the crystal. For the Tevatron, with crystal at 5σ, the probability of deflection beyond a 5.5σ aperture is 0.5% per every encounter with a crystal. As a 5 mm Si crystal is just ~1% of a proton nuclear interaction length, the number of encounters in the accelerator ring can potentially be very high, and the accumulated probability of volume capture quite significant.

In a circular accelerator, the scattered particles continue the circulation in the ring and encounter the crystal again and again on later turns. The particle distribution modifies due to betatron oscillations in the ring and due to scattering on every encounter with the crystal. For crystal orientation within the plateau in Figs. 1-2, the conditions for coherent scattering and volume capture take place on every encounter with the crystal. For random orientation, this never happens. Respectively, the particle amplitude in accelerator grows faster if coherent processes contribute strongly to the overall scattering. By every encounter with a crystal, beam emittance grows by about $\beta(\Delta\theta)^2$, where $\beta$ is accelerator beta function and $\Delta\theta$ is the scattering angle in crystal. With the O-shaped crystal used in the experiments and simulations, we find that beam emittance grows faster by a factor of 3 in RHIC case, and factor of 5 in the Tevatron case, in the plateau region compared to random alignment. This difference in beam dynamics on the phase space with a crystal under conditions of coherent scattering leads to a faster particle loss on the secondary elements of accelerator. As a result, the particle loss (nuclear interactions) in crystal is reduced, as part of the loss now goes to different elements in the ring.

## 4. Predictions for the Tevatron

In order to check our understanding and make further predictions, we simulated the ongoing Tevatron collimation experiment. More details of the used settings are in ref. [17,30]. We used the recent edition of Tevatron accelerator lattice [31]. In the bending plane, the accelerator functions were β=60.205 m and α=-0.216 at the crystal location. The O-shaped crystal was placed at 5σ and served as a primary element in collimation scheme. The secondary collimator was placed 31.5 m downstream, at 5.5σ. Particle tracking in the Tevatron lattice is done with linear transfer matrices. Each particle was allowed to make an unlimited number of turns in the ring and of encounters with the crystal until a particle either undergoes a nuclear interaction in the crystal or hits the secondary collimator (either because of a bending effect in the channeling crystal or because of the scattering events). A non-channeling amorphous layer 2 micron thick was assumed on the crystal surface due to its irregularity at a micron level.

The simulated nuclear interaction rate in the O-shaped crystal is shown in Fig. 2 [24] together with the experimental data. The remarkable dip on the plot, nearly 95% down from the rate observed at random orientation, is due to channeling with a high efficiency in the environment of the Tevatron collimation experiment. One can expect about an order of magnitude reduction of machine-related backgrounds in the collider detector if crystal components are installed in the collimation systems of collider. The dip is ~30 μrad wide, which is greater than 2 critical angles because here channeling is essentially multipass, multiturn effect. Protons encounter crystal several times, scatter, go on circulating in the ring, and then get channeled on some later encounter. This scattering contributes sizably to the width of the peak. A distinct feature in Fig. 2 is the plateau on the plot, with a 50% reduction in the rate that is in good quantitative agreement with the experiment.

It is interesting to compare the relative contributions of volume capture and volume reflection to the plateau effect. From all the protons intercepted by the secondary collimator in the Tevatron, roughly 80% are peaked at the edge while ~20% have a long flat distribution with big impact parameters. These 20% of the aperture loss are "volume-captured" protons while the 80% can be called "volume-reflected" particles. Their origin can be roughly understood also in the following way. A typical number of proton encounters with the crystal before a proton loss was ~20. With the ~0.5% probability (per encounter) of channeling a proton to the 5.5σ aperture, the overall probability of volume-capture channeling is ~10%, thus producing this long flat tail of big impact parameters. Volume-reflected protons diffuse to the aperture. Some of them are lost at its edge while others die in crystal.

So, basic explanation to the plateau effect in collimation is a strong coherent contribution to the overall scattering, affecting the particle loss pattern along the accelerator. The increase in the crystal scattering angle from coherent effects makes a beam diffusion to increase several-fold. This diffusion and the presence of the secondary aperture make the effect of plateau. However, the volume-capture contribution (a non-diffusion term) is not negligible, making the overall kinetic picture more complicated.

New experimental tests would be decisive for the theory. A new bent crystal (strip-type deflector produced in IHEP) is now replacing the O-shaped crystal in the Tevatron [24]. The new crystal is Si (110), 3 mm along the beam, 0.15 mrad bent. Before the measurements started with it, we present our predictions in Fig. 4. The rate suppression of 65% at plateau, 0.15 mrad wide, is expected with the new crystal in the same setup. The channeling dip at a crystal best alignment is expected somewhat deeper than with the old crystal, about 96%. The most probable reflection angle for 980 GeV proton in this crystal is –8.5 μrad, which is 1.7 times the critical angle.

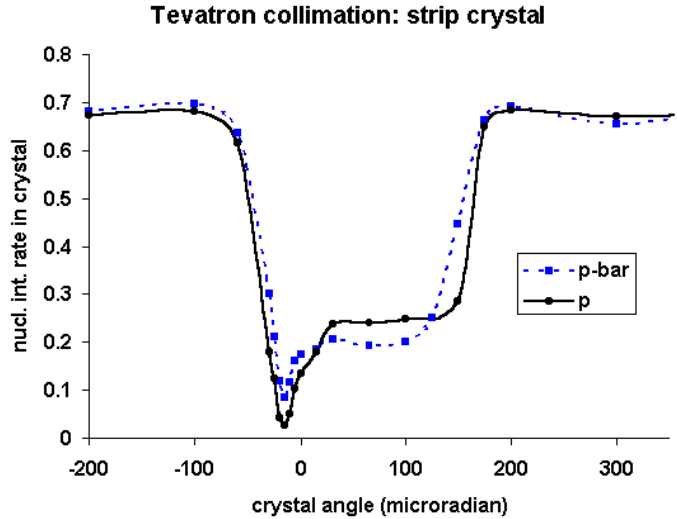

**Figure 4** The predicted crystal nuclear interaction rate for protons and antiprotons in the Tevatron with the "strip" crystal.

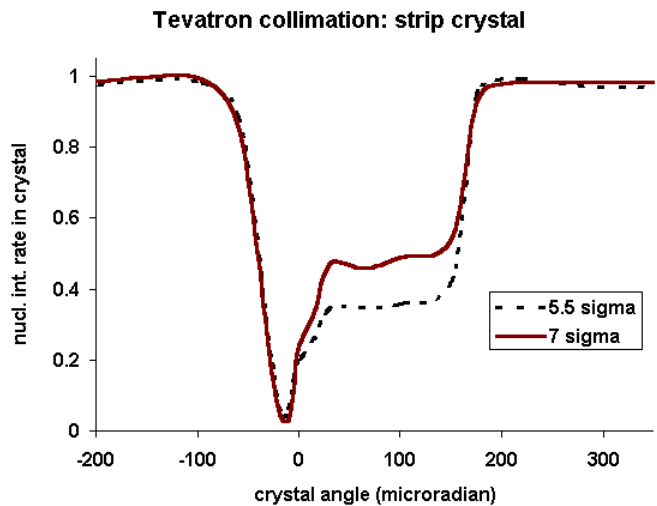

**Figure 5** The predicted crystal nuclear interaction rate for protons in the Tevatron with collimator set at 5.5 and 7 sigma.

In order to study the behavior of the plateau effect in different collimation geometry, we show another prediction in Fig. 5. Here the secondary collimator is placed at 7σ while the crystal is kept at 5σ. Thus, the horizontal offset is increased from 0.5 to 2σ. In this setup, the predicted rate suppression is 50% at plateau, and the channeling dip at crystal best alignment is 97%.

## 5. Negative particles

A very interesting question is how this effect applies to particles of negative charge. Volume reflection is known also for negative particles, although its expected magnitude is much lower than for positive ones [27]. At the Tevatron, it is possible to study crystal physics in the same setup also with particles of negative charge, antiprotons. For channeling phenomenon, the sign of particle charge plays a critical role. For instance, dechanneling is very strong for negative particles; therefore bent crystal channeling made only negligible bending effect for negatives [32]. However, our theory predicts that a crystal collimation effect for antiprotons is about as strong as one for protons, Fig. 4. We suggest to check this prediction in the Tevatron with antiprotons, as this appears critical for understanding of crystal collimation mechanism and, crucially, it also opens a principle way to efficient crystal steering of negatively charged particles at accelerators. In a single encounter, Fig. 6, the *mean* exit angle for antiprotons is just -1.6 μrad while for protons it is -7.5 μrad (and most probable exit angle is respectively -4 and -9 μrad). However, in our theory the effect is run by the *mean square* exit angle, and we find the r.m.s. angle about the same for protons and antiprotons, ~12 μrad. At plateau, the rate is suppressed by ~70% for antiprotons.

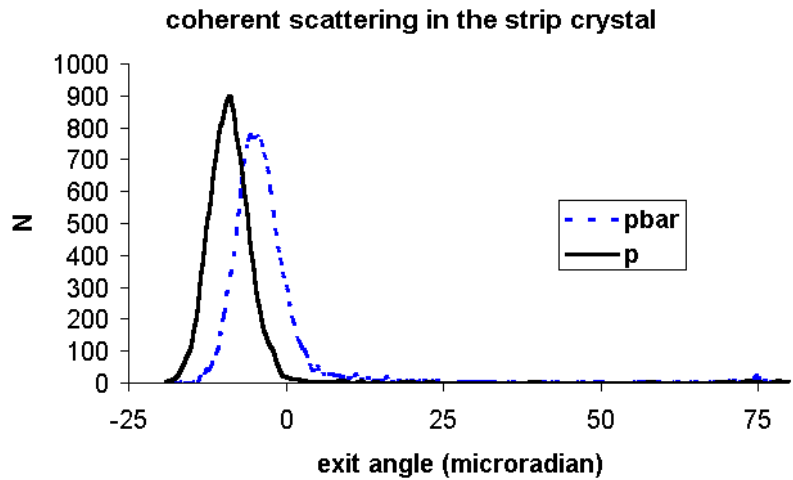

**Figure 6** Exit angular distributions for 980 GeV proton and antiproton in the "strip" crystal.

Usually, plain channeling is not useful for bending of negative particles because of very strong dechanneling [32]. However, it was already noticed in 1998 [33] that at TeV energies in case of collimation (because of small bending angles involved) the negative particles could be efficiently channeled, e.g. for collimation of 2-TeV negative muons [33]. In the Tevatron case we find a significant channeling effect for 1 TeV antiprotons because here a bending effect of merely ~25 μrad is significant. At the dip in Fig. 5, the rate suppression for antiprotons is 87%.

We say again that the mean angular kick for antiprotons is factor of 5 smaller than for protons, but the collimation effect at the plateau for antiprotons is at least as strong as for protons. This clearly contradicts the intuitive view that the mean exit angle different from zero (i.e. literally reflection) is responsible for the plateau effect in crystal collimation. The Tevatron experiment could answer whether our theory is true.

## 6. Crystal model test in CERN SPS H8 experiment

The effect of volume reflection is observed in crystal collimation indirectly, through increased beam diffusion. An experiment on a direct measurement of volume reflection angle and reflection efficiency has been in progress at CERN SPS [34] making use of H8 external 400 GeV proton micro-beam line with unique possibilities for crystal tests. This experiment allows another check of the theory. Our model has predicted [35] for the SPS experiment a reflection with most probable angle of 13 μrad and efficiency of 96%; see more details in ref. [36].

## 7. Bent crystal as a smart material

Strong scattering over a short length means that the effective scattering length is strongly reduced compared to a radiation length $L_R$ of an amorphous material. Some of the Si crystals produced at PNPI [37] and tested at IHEP were just 0.3 mm along the beam, with bending of some hundred μrad. In such a short crystal with bending of 20 μrad, a 7 TeV proton obtains ~2.2 μrad r.m.s. angle of coherent scattering, according to simulation. This is equivalent to a "radiation length" of just 0.25 mm, a factor of 400 shorter than $L_R$ in amorphous Silicon. According to simulation, the effective "radiation length" in a bent W(110) crystal can be just ~10 μm. At the same time, the length of inelastic nuclear interaction $L_N$ reduces insignificantly. As a result, the $L_N/L_R$ ratio can reach ~10000 in this "smart material" while in amorphous materials it is just ~5 to 30. In accelerator ring, such a smart target would be an extremely efficient scatterer.

## 8. Summary

Crystal collimation studies, besides promising a very high efficiency of the technique in colliders, reveal a new interesting physics of beam scattering off the coherent field of bent crystal atomic planes. This coherent scattering causes a perturbation (diffusion) of beam in the conditions of crystal collimation experiments at RHIC and Tevatron and is observed as a strong factor affecting particle loss in the accelerator ring. This physics is essential for crystal applications in colliders. However, it exists also at MeV energies as shown in another recent study [38]. Possible applications of volume reflection at IHEP are considered in ref. [39].

Our computer model is in good quantitative agreement with RHIC and Tevatron data. The theory predicts that the factor behind the new coherent effects in crystal collimation is a strong increase in the *mean square* angle of a particle scattered off the coherent field of a bent crystal. We show that this opens way for efficient bent crystal steering of negative particles, which may help, e.g., with collimation of antiprotons and muons.

Currently, accelerator physicists use a choice of amorphous scattering materials from low-Z like carbon to high-Z like tungsten. Crystals can be realized also as smart materials with effective scattering length three orders of magnitude shorter than a radiation length of amorphous material. Nanotechnology offers more opportunity to construct smart materials for beam steering [40]. Crystal technique could improve the efficiency of scattering and collimation by orders of magnitude.